\documentclass[reprint,amsmath,amssymb,aps]{revtex4-2}
\usepackage{graphicx}
\usepackage{dcolumn}
\usepackage{gensymb}
\usepackage{hyperref}

\usepackage{adjustbox}
\usepackage{ulem}
\usepackage{cancel}

\usepackage[usenames,dvipsnames]{xcolor}

\begin{document}

\title{An optimized recipe for making giant bubble}%

\author{Marina Pasquet$^{1,*}$, Laura Wallon$^{1,*}$,  Pierre-Yves Fusier$^{2}$, Frédéric Restagno$^1$, and Emmanuelle Rio$^1$, }%
\affiliation{$^{1}$ Universit\'e Paris-Saclay, CNRS, Laboratoire de Physique des Solides, 91405, Orsay, France. \\
}
\affiliation{$^{2}$ Slash bubbles, rue de Versailles, 91300 Massy, France\\
}

\date{September 7, 2022}%

\begin{abstract}
Big bubbles are largely used in physics laboratories to study 2D turbulence, surface wavers, fundamental properties of soap systems\dots On a more artistic point of view, blowing big bubbles is part of many artistic shows. Both communities usually wan to get reasonably stable foam films. The purpose of this article is to propose the main physical ingredients allowing to identify a good recipe for making stable films and bubbles. 
We propose controlled experiments, to measure both the easiness to generate a bubble and its stability for different stabilizing solutions, which we choose by adding one by one the ingredients contained in an artist’s recipe. 
The main results are that (i) the surfactant concentration must be not too high (ii) the solution must contain some long flexible polymer chains to allow an easy bubble generation and (iii) the addition of glycerol allows a better bubble stability by avoiding evaporation. 
We finally propose an efficient recipe, which takes into account all these considerations.
\end{abstract}

\maketitle

\section{Introduction}

Blowing bubbles is an experience enjoyed by young and old alike. 
In the 17th century, both Robert Hooke \cite{hooke28holes} and later Isaac Newton \cite{newton1952opticks} have been excited by the beauty of soap bubbles.  
After Newton, a number of scientists performed numerous and curious
experiments with soap bubbles and films \cite{boys1888li,isenberg1992science}. 
More recently, soap films and bubbles have been the object of numerous scientific studies. 
Some of them focus on their stability but we can identify many research problems, in which the foam film or bubble rupture only marks the unwanted end of the experiment.
This is the case for the studies focused on film and bubbles structure \cite{WEAIRE200820,drenckhan2015structure} as well as on 2D turbulence \cite{Kellay_2002} or more generally on hydrodynamics in soap films \cite{chomaz1990soap,couder1989hydrodynamics,Kellay_POF_2017}.
In such studies, it reveals very important to identify the best recipe to obtain stable foam films and bubbles and allow an investigation during a reasonable time.

\begin{figure}
\centering
\includegraphics[width=1\linewidth]{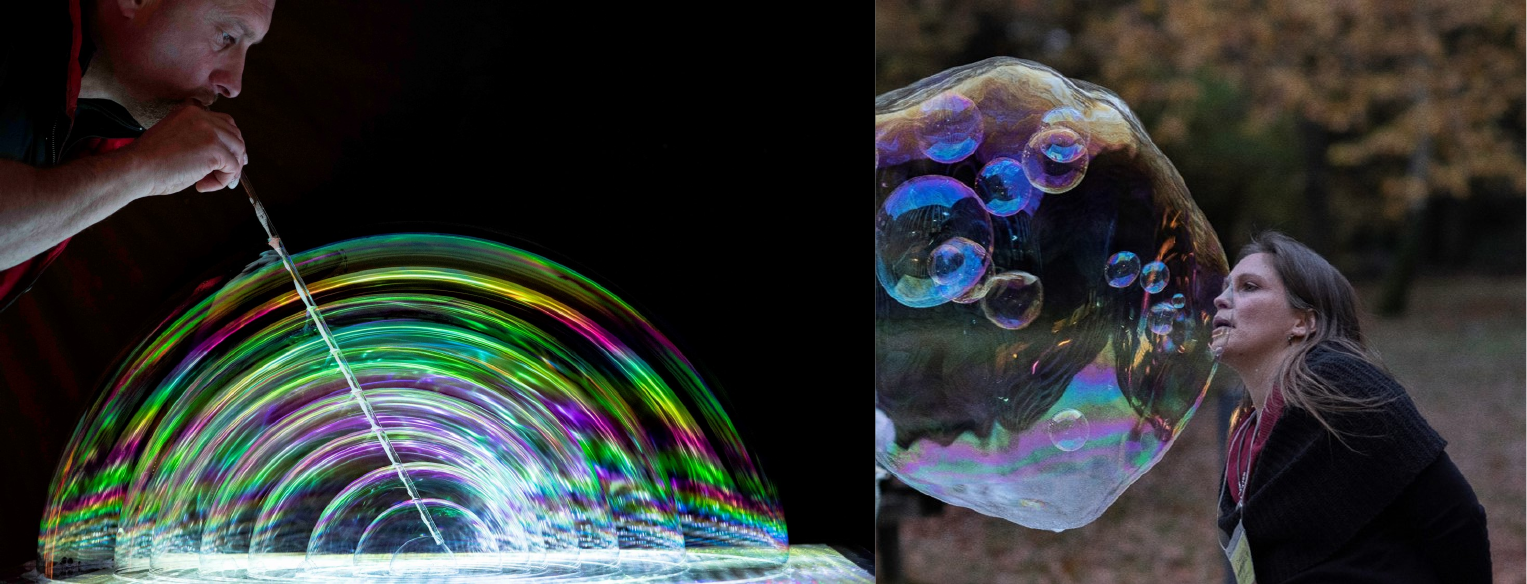}
\caption{(a) Slash bubbles\footnote{https:$//$www.slashbubblesparis.com$/$}  an artist, bubble trainer, blowing big concentric bubbles. (b) Miss bubble bliss\footnote{http:$//$missbubblebliss.at$/$} blowing bubbles into a giant bubble. Pictures taken by Serge Guichard.}
\label{fig:PierreYves_Stephanie}
\end{figure}

There is an important know-how concerning the best recipes not only in the scientific community but also in the artistic one as shown by the performance they can reach (Fig. \ref{fig:PierreYves_Stephanie}).
For example, the record for the largest outdoor free floating soap bubble was achieved by Gary Pearlman (USA) in June 2015 with a bubble of 96.27 m$^2$ \cite{webPealman}. 
This knowledge comes from years of trials and errors by the artistic community \cite{Soapwiki} but a comprehensive understanding is lacking. 
Comparing the different proposed recipes allows drawing some first conclusions. 
The main ingredients used are commercial detergent dissolved in water together with the addition of small quantities of polymers, glycerol as well as some various additives. 
Recently, Frazier \textit{et al.} \cite{frazier2020make} have explored the role of high long-chain polymers in bubbles solutions and they have evidenced the role of extensional rheology of these dilute polymer solutions as an important factor in creating the films.

In this article, we propose to build on the know-how of the artist community to identify the important ingredients contained in their best recipes and perform controlled experiments to quantify the effect of each ingredient and rationalize the identification of a good recipe, which can be used and adapted at will by researchers. 

\section{Material and methods}

A good starting point to obtain a good recipe is to start by reading \cite{Soapwiki} and by discussing with artists, who have already developed an extensive know-how to make amazing bubbles. We identified the main ingredients in their recipe and proposed a classification to help us choose a typical recipe. 
Every recipe contains the following ingredients (the concentrations are given in volume percent):

\begin{itemize}
    \item Soap bubbles need soap. 
    The main ingredient is thus dishwashing liquid. 
    Almost every recipe proposes to use Fairy or Dreft (dishwashing liquids sold by Procter \& Gamble). Their formulation is made of Sodium lauryl polyoxyethylene ether sulfate (15-30~\%), ethanol (1-5~\%), amines, C10-19 alkyldimethyl, N-oxides (5-10~\%) and Alcohols, C9-11, ethoxylated (1-5~\%). The typical concentration used in solution is around a few percents. 

    \item Long aqueous polymers are often added to the solution. 
    Depending on the recipe, it can be J-Lube (Jorgensen Labs), which is composed of 25 \% polydisperse Poly Ethylen Oxyde PEO
(up to 8 × 10$^6$ g/mol) and 75 \% sucrose. It can more simply be Guar gum (a food additive). 
    More rarely, some artists use polymers such as cellulose gum CMC, hydroxyethyl cellulose HEC, hypromellose HMPC, xanthan gum, polyacrylamide PAM.
    These have in common to be long polymer chains soluble in water. 
    Their concentration in solutions is small, in general, around or less than 0.1 \%. 

    \item     Glycerol is often added. Some artists use it to facilitate the mixing of ingredients and therefore do not control its concentration. Others add it at controlled concentrations ranging from a few percent to a few tens of percent.

    \item Various additives can also be found in the recipes such as citric acid, yeast\dots 
    They are added in very small quantity (less than 0.1 \%) and often used to adjust the pH of the solution. In their absence the pH of the solutions is usually around 9.

\end{itemize}

In the following, we propose to focus on a relatively simple but quite efficient recipe used daily by one of the authors of this article, who is a bubble trainer. 
We will first test the best dishwashing liquid concentration. Then we will add one by one the other ingredients to observe how they affect the quality of the recipe. 

We worked with Fairy with an anionic surfactant concentration between 15 and 30 \%, with J-Lube and Guar gum (G4129-250G purchased from Sigma), with glycerol (purity $\leq$ 99.5 \% purchased from VWR), with citric acid (Prolabo), with yeast (Alsa) and the water used was ultrapure water (resistivity = 18.2 M$\Omega \cdot $cm).

\section{Experimental set-up}

In the following we propose two different experiments, one to test the efficiency of the recipe when you try to blow bubbles (section \ref{Part:BlowingBubbles}) and the second one to test the stability of the obtained bubbles (section \ref{Part:BubblesStability}). We will also describe the protocol used to measure the elongational viscosity of solutions containing Guar gum (section \ref{Part:BubblesStability}).

\subsection{Blowing bubbles} \label{Part:BlowingBubbles}

\begin{figure}
\centering
\includegraphics[width=1\linewidth]{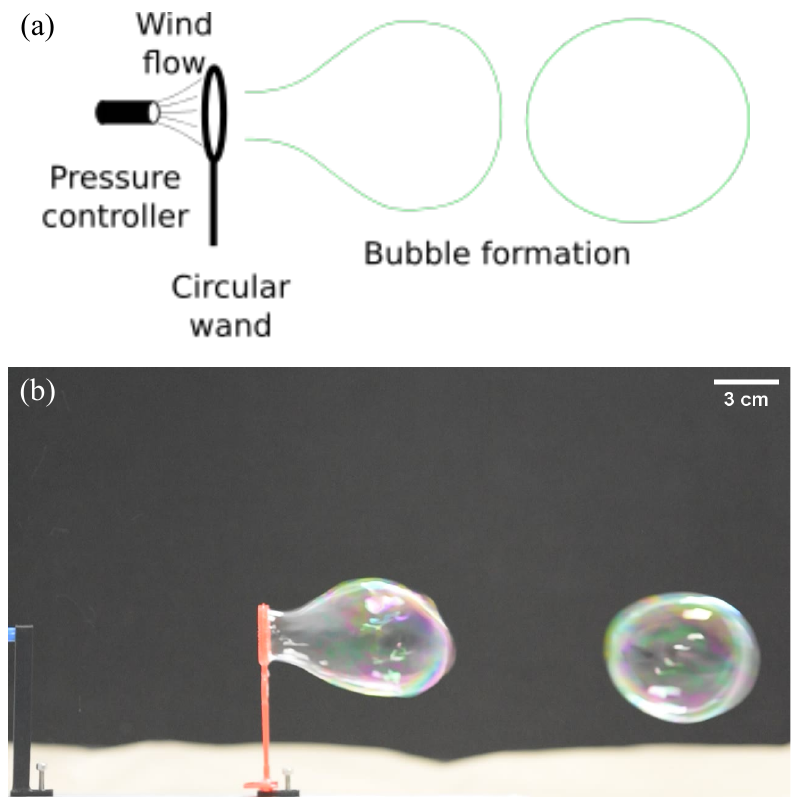}
\caption{(a) Scheme of the blowing bubble experiment. A pressure controller blows air at constant velocity (22 m/s) in a wand, which has been plunged in the soapy solution. A camera makes it possible to film the generation
of bubbles and count them. (b) Photograph of the bubble-blowing experience. A centimetric, vertical soap film
is placed in front of an airflow generator.}
\label{fig:SetUpGeneration}
\end{figure}

Our setup is schematized in Fig. \ref{fig:SetUpGeneration}. 
To observe the generation of bubbles, we set up a device, in which a soap film is formed on a child's toy (with a diameter of 2.7~cm) and placed vertically in front of an airflow generator at 22~m/s.
A blowing velocity, of a few tens of meters/second allows to be in a regime, where many bubbles can be created, whose size is fixed by the size of the wand \cite{hamlett2021blowing,salkin2016generating}.

A camera (Nikon D7200) with an objective (Nikon AF Micro 200 mm f/4.0D) is placed in front of the device and allows making side views (Movie M1). 
For each film, the number of bubbles made before the soap film breaks is counted manually using the recorded movies of the experiments. 
We calculated the probability of failure, \textit{i.e}. the ratio between the number of times no bubbles were made before the film was bursting and the number of times bubbles was made. 
The experiment is done 50 times for each solution.

We can notice that this experiment does not completely mimic the generation of a bubble by blowing on a soap film "by hand" but could clearly mimic the generation of bubbles by commercial apparatus or automated lab experiments. 

\subsection{Bubbles stability} \label{Part:BubblesStability}
\begin{figure}
\centering
\includegraphics[width=1\linewidth]{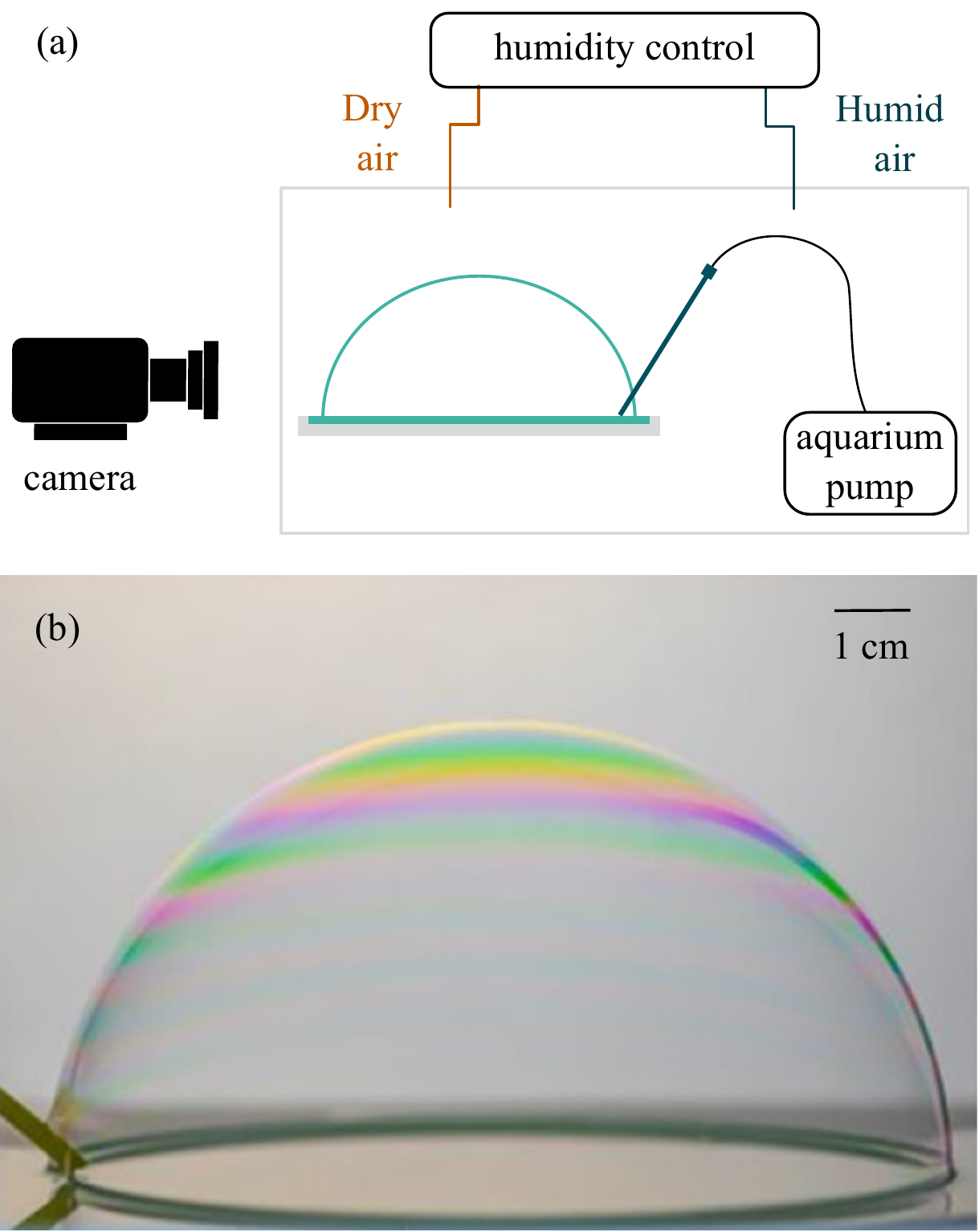}
\caption{(a) Schematic of the automated experiment to generate large surface bubbles. The bubbles
of 5.85 cm in radius are created by blowing with a straw below the surface in a humidity-controlled box. A LabView program allows direct visual detection of the presence of a bubble: the generation is thus automated in order to measure their lifetime on about
fifty bubbles for each experiment. (b) Example of a bubble obtained with this device. Photo credits: Serge Guichard.
}
\label{fig:ManipGrosseBulle}
\end{figure}

Our second setup is schematized in Fig. \ref{fig:ManipGrosseBulle} (a) and allows us to make automated measurements. 
A glovebox type enclosure of size 57.5~cm~$\times$~43~cm~$\times$~32~cm, in which the humidity is controlled thanks to a LabView program, is the location of the measurements.
Humidity is controlled by the same principle as proposed by Boulogne \cite{Boulogne2019}, namely that a relay controls two aquarium pumps allowing blowing either dry or humid air depending on the ambient conditions. Humidity and temperature are measured and recorded over time, using two sensors (SHT25 purchased from RS).
All our experiments were performed at an atmospheric humidity $\text{RH} = 60 \pm 5 \%$. 
Bubbles are generated by blowing through a straw under the surface of a bath in a PTFE tank of 18 cm diameter and 5 mm depth (Fig. \ref{fig:ManipGrosseBulle}). 
The solution to be studied should fill the bath to form a meniscus. 
To control the bubble volume, air is injected at constant flow using an aquarium pump (Tetra 400) for a fixed time of 20~seconds, controlled with a relay. 
The radius of the obtained bubbles is 5.85~cm. The size is sufficiently small to consider that the bubbles are not affected by gravity \cite{cohen2017shape}.

To measure the lifetime of the bubbles in an automated way, we detect their presence or absence with the use of images analyzed over time and obtained with a camera (acA1300-60gm Basler).
In order to have a good automated detection, it is important to adjust the lighting of the bubble. 
We use an LED lamp (Gdansk, 29.5~cm~$\times$ 29.5~cm purchased from Leroy Merlin) and a diffusive plate in order not to saturate the camera, which is in front of the illuminating system.

Before each measurement, a reference image must be made which will be used to detect the presence of the bubble by image subtraction. 
If the difference between the two photographs is greater than a certain threshold which can be set by the program, it is considered that a bubble is present. 
The chronometer starts just after the air injection is stopped.
When a bubble is no longer detected on 10 consecutive images, the program stops the chronometer and starts the injection of the next bubble. 
The program also measures bubble size in real time  (Movie M2), by detecting the bubble contour, to prevent potential problems and to check the reproducibility of the measurements.
The lifetime of each bubble is then recorded as well as the radius of the bubbles, the humidity and temperature in the box during the measurement. 
The lifetime is measured at least 50 times for each solution.

\subsection{Measuring the elongational viscosity} 
\label{sec:ElongVisc}
The elongational viscosity can be measured using a DoS (Dripping-onto-Substrate) rheometry protocol. 
We have reproduced the one used by Dinic \textit{et al.}  \cite{Dinic2017}. 
This method consists in visualizing with a high-speed camera (Photron Fastcam SA3 with a telecentric lens at 6000 fps) and analyzing the capillary thinning of a filament that forms when a drop of the studied solutions falls at a controlled rate ($Q = 0.02$~mL/min) on a microscope slide (Fig. \ref{fig:ElongationalViscosity} inset). 
The flow rate is controlled with a syringe pump and the light source (Phlox lamp) is placed in transmission with a diffuser. 
The diameter of the needle at the end of which the drop is formed is 1.54~mm, it is located at about 5~mm from the microscope slide.

\section{Results}

\begin{figure}
\centering
\includegraphics[width=1\linewidth]{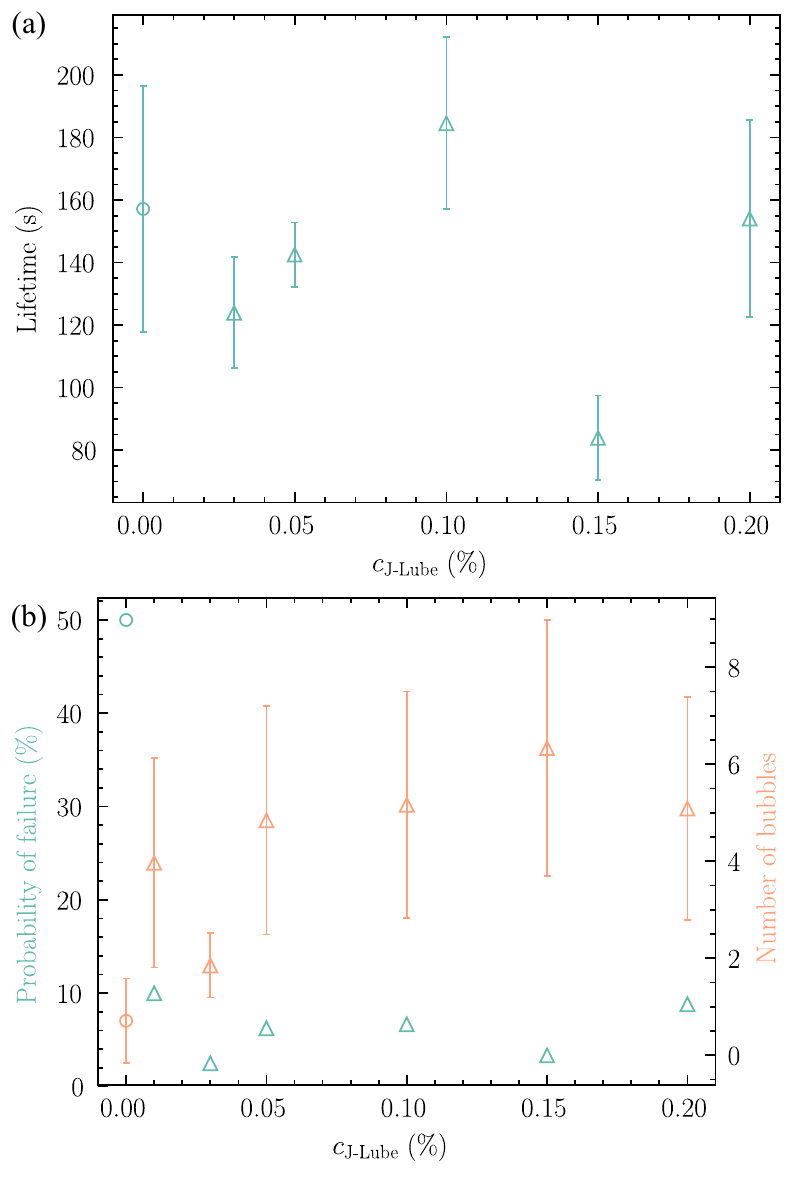}
\caption{(a) Average bubble lifetime measured as a function of the concentration of J-Lube in the solution (composed with water and Fairy at a concentration of 4 \%), using the device shown in Fig. \ref{fig:ManipGrosseBulle}. The error bars correspond to the standard deviation measured on about 50 bubbles. (b) For the same solutions, this graph shows the evolution of the failure probability (left axis) and the number of blown bubbles (right axis) obtained with the setup presented in Fig. \ref{fig:SetUpGeneration}. The error bars correspond to the standard deviation measured on about 50 trials.}
\label{fig:EffetConcentrationJLube}
\end{figure}

The results obtained for each solution, with the two experiments presented in the previous section, are plotted in Fig. \ref{fig:EffetConcentrationFairy} and  \ref{fig:EffetConcentrationGuar}-\ref{fig:EffectAdditives}. 
All these figures have two parts.  In Fig. (a), we plotted the average bubble lifetime for each solution and the error bars are given by the standard deviation obtained on the measurements.  
In Fig. (b), we plotted the probability of failure \textit{i.e} the probability to have zero bubbles by blowing in a film (right axis) and the number of bubbles created by blowing in the film (left axis). 

\subsection{Effect of the dishwashing liquid concentration}

\begin{figure}
\centering
\includegraphics[width=1\linewidth]{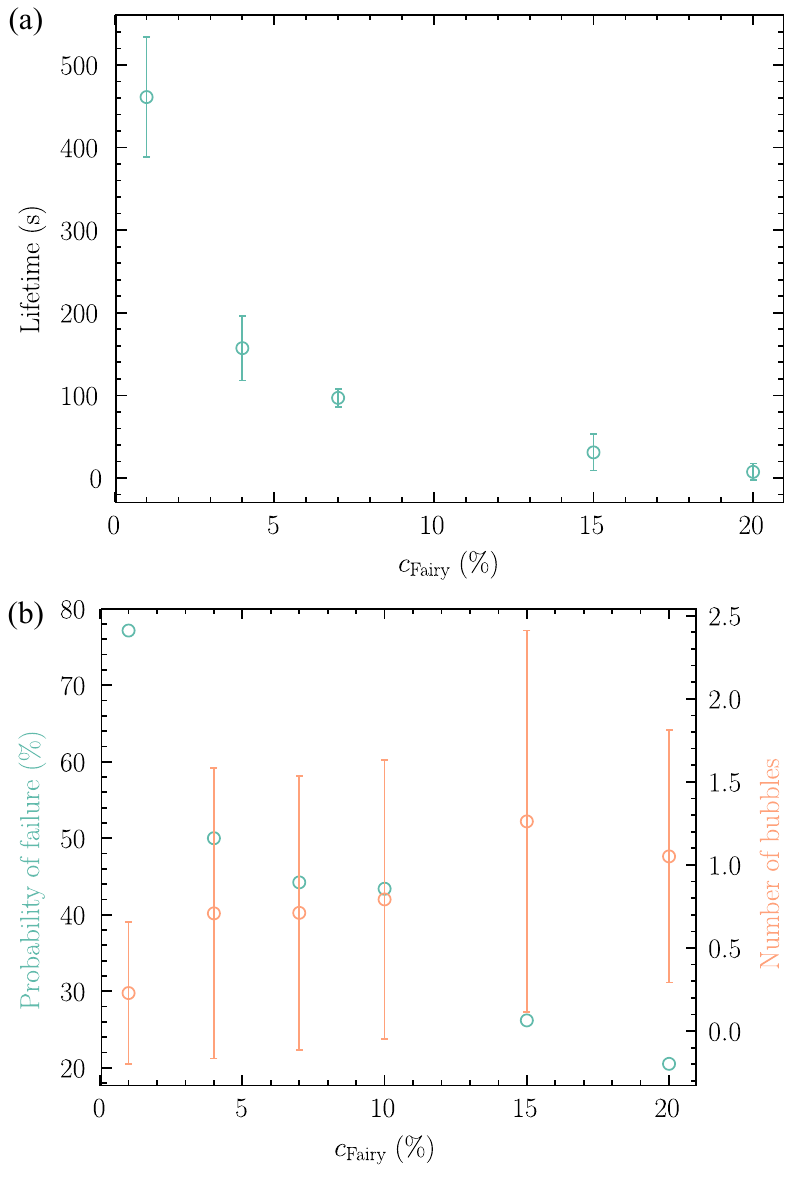}
\caption{(a) Average bubble lifetime measured as a function of the concentration of Fairy in the solution (in this case composed only of water and Fairy), using the device shown in Fig. \ref{fig:ManipGrosseBulle}.  The error bars correspond to the standard deviation measured on about 50 bubbles. (b) For the same solutions, this graph shows the evolution of the failure probability (left axis) and the number of blown bubbles (right axis) obtained with the setup presented in Fig. \ref{fig:SetUpGeneration}. The error bars correspond to the standard deviation measured on about 50 trials.}
\label{fig:EffetConcentrationFairy}
\end{figure}

\paragraph{Results.} The results concerning the impact of the concentration in dishwashing liquid (Fairy) dissolved in water are presented in Fig. \ref{fig:EffetConcentrationFairy}. 
Concerning the stability, we observe that the more the Fairy concentration decreases, the more the bubbles are stable with a ratio around 50 between the lifetime obtained at 1 \% and the one obtained at 20 \%. 
There is probably an optimum since we know that the stability will be very low for pure water bubbles. 

However, it is very difficult to decrease the concentration under 1 \%. We can see in Fig. \ref{fig:EffetConcentrationFairy} (b) that the probability of failure increases when the concentration of Fairy decreases and becomes equal to 75 \% for a Fairy concentration equal to 1 \%: for a low soap concentration, it becomes highly probable to be unable to blow any bubble.
Thus, concerning the ease of bubble generation, we observe the higher the concentration, the more bubbles are generated by blowing. 
This probably explains the empirical choice of a concentration of 4 \% made by the artists and proposed in a scientific study \cite{ballet2006giant}.
It is indeed a compromise between the stability of the bubbles and the possibility to generate them. In the following, we will concentrate on solutions containing 4 \% of diswashing liquid.

\begin{figure}
\centering
\includegraphics[width=1\linewidth]{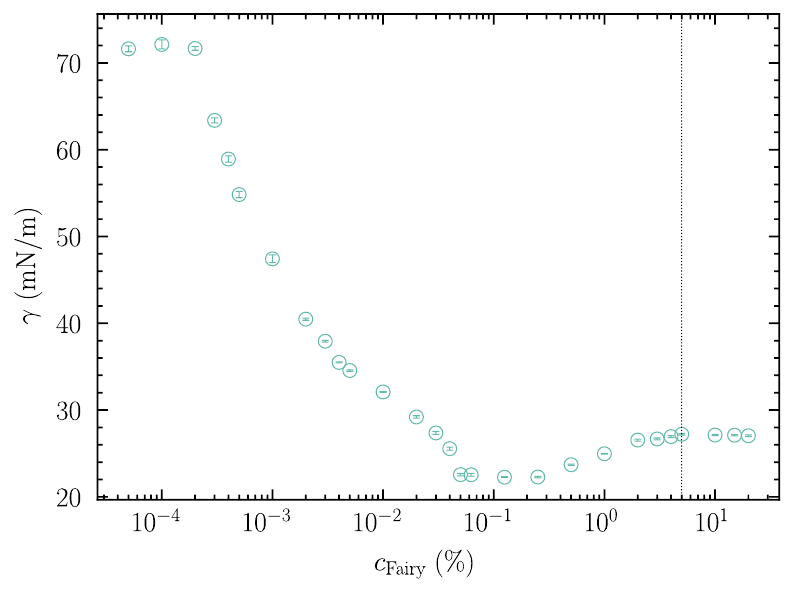}
\caption{Measure of the surface tension of solution of Fairy in water as a function of the concentration in Fairy in \%. The plateau at high concentration corresponds to the existence of micelles in solutions. One of the latter is schematized in the figure. The vertical line corresponds to the concentration often used in the solutions $c_{\text{Fairy}} = 4 \ \%$.}
\label{fig:Marangoni}
\end{figure}

\paragraph{Discussion.} Let us try to rationalize the observation that it is necessary to add soap to stabilize bubbles, with a small enough concentration. For that,  we measured the surface tension as a function of the concentration in dishwashing liquid for the Fairy, which has been used during this study. The result is plotted in Fig. \ref{fig:Marangoni}.
We observe a very classical shape, expected for pure surfactant solutions \cite{Bergeron1995}: the surface tension decreases when the surfactants concentration increases. After a certain concentration of surfactants, the surface tension remains constant which is a sign of the formation of micelles in the solutions.  The minimum can be attributed to the presence of impurities \cite{Lin1999,mysels1986surface}.

To be able to create a foam film, before blowing, the soap film is pulled out of the soapy solution. It is only the gradient of surfactants between the bottom of the film and the top of the film which holds the film  \cite{DeGennes_2001,couder1989hydrodynamics,Lucassen1981}. 
A surface tension gradient and thus a surfactant surface concentration gradient is necessary to stabilize a soap film. 
If the bulk surfactant concentration is too high, the surfactant can repopulate very easily the interface which prevents the establishment of surface tension gradients.
The concentration at which our experiments are performed, 4 \% of Fairy is materialized in Fig. \ref{fig:Marangoni}(a) by the vertical dashed line. 
It is at the beginning of the Plateau, above the cmc but not much, in agreement with the previous explanation.

\subsection{Effect of the addition of lubricant}

\begin{figure}
\centering
\includegraphics[width=1\linewidth]{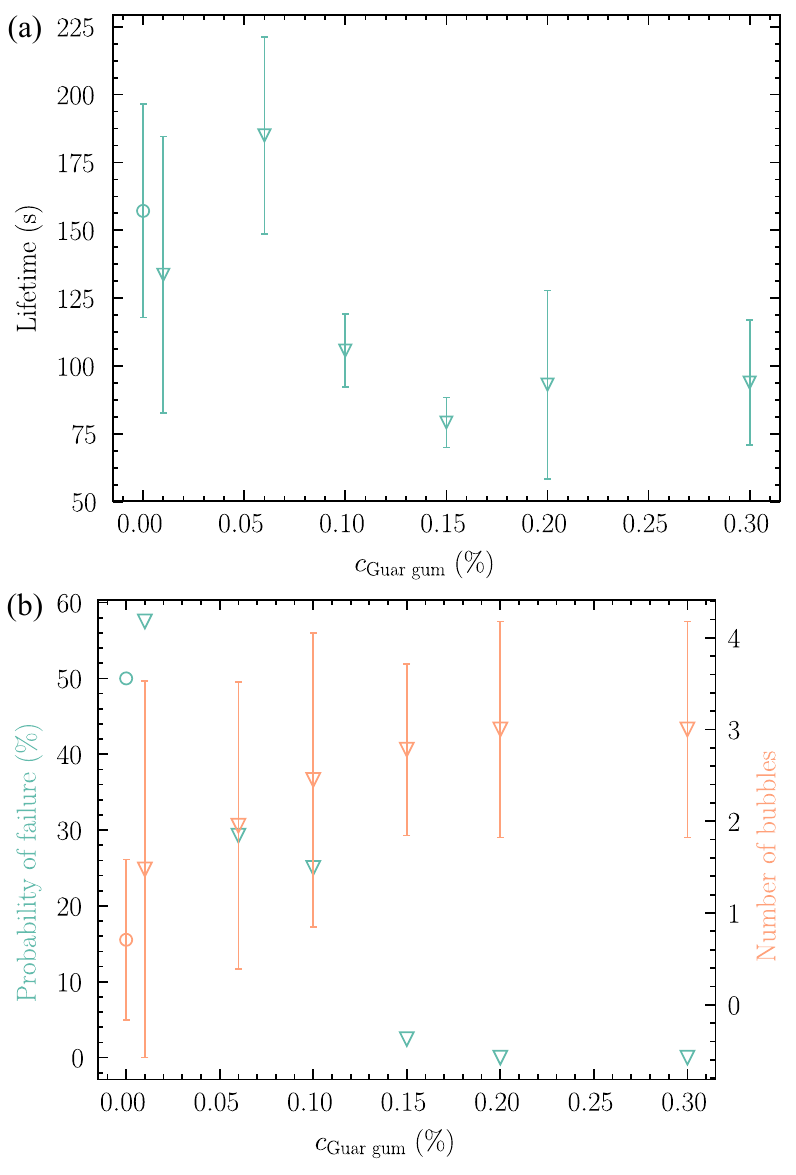}
\caption{(a) Average bubble lifetime measured as a function of the concentration of Guar gum in the solution (composed with water and Fairy at a concentration of 4 \%), using the device shown in Fig. \ref{fig:ManipGrosseBulle}. The error bars correspond to the standard deviation measured on about 50 bubbles. (b) For the same solutions, this graph shows the evolution of the failure probability (left axis) and the number of blown bubbles (right axis) obtained with the setup presented in Fig. \ref{fig:SetUpGeneration}. The error bars correspond to the standard deviation measured on about 50 trials.}
\label{fig:EffetConcentrationGuar}
\end{figure}

\paragraph{Results.} We then looked at the effect of Guar gum concentration (Fig. \ref{fig:EffetConcentrationGuar}) by adding Guar gum in a solution of Fairy with a concentration equal to 4~\%. Due to the dispersion of the experimental data, it is almost impossible to measure a noticeable effect of the Guar gum concentration, in the range of studied concentrations, on the soap bubbles lifetime.
On the other hand, it is much easier to generate bubbles in the presence of Guar gum. Indeed, adding some Guar gum to the solutions makes  the probability of generation failure practically zero above a concentration of 0.15 \%. Another remarkable result is the number of bubbles which have been generated: we go from 0-1 bubbles without Guar gum to 2-4.
Similar results have been obtained by adding J-Lube to the dishwashing liquid solution (Fig. \ref{fig:EffetConcentrationGuar}).

\paragraph{Discussion.} An important parameter, which is a lot affected by the presence of a long chain polymer like Guar gum or PEO, which is present in J-Lube, is the elongational viscosity $\eta_e$. 

\begin{figure}
\centering
\includegraphics[width=1\linewidth]{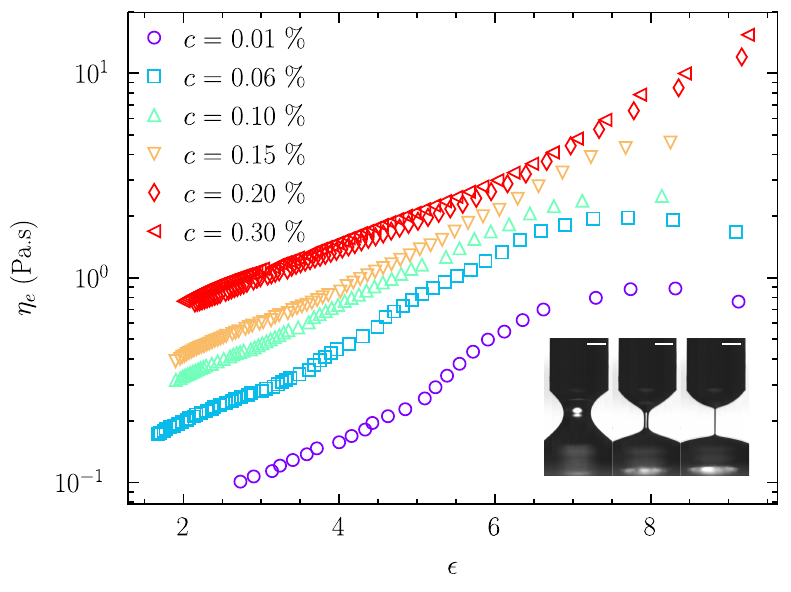}
\caption{Evolution of the elongational viscosity $\eta_e$ as a function of the Hencky strain $\epsilon$ for different Guar gum concentrations $c$ in the solutions (composed with water and Fairy at a concentration of 4 \%), using the DoS protocol described in \cite{Dinic2017}. The snapshots shown are for thinning neck of a solution
with $c = 0.1 \ \%$. The scale bar is 0.5~mm.}
\label{fig:ElongationalViscosity}
\end{figure}

We made measurements with solutions stabilized with Guar gum (solutions identical to those in Fig. \ref{fig:EffetConcentrationGuar}), as the properties of this polymer are more controlled than commercially available J-Lube powders, in the perspective of using it in laboratories. 
The protocol is presented in section \ref{sec:ElongVisc}.
The results are given in Fig. \ref{fig:ElongationalViscosity} as a function of the Hencky strain $\epsilon$:
\begin{equation}
\epsilon  = 2 \ln \left( \dfrac{\mathcal{R}_0 }{\mathcal{R}} \right)   
\end{equation}
where $\mathcal{R}$ denotes the radius of the neck which evolves with time and $\mathcal{R}_0$ its initial value.
In our situation, we start with a film of the order of a 5 $\mu$m (typical thickness of a soap film obtained by pulling a frame out of a soapy solution) that we stretch to thicknesses of the order of 500 nm (soap films exhibiting bright colors). 
This corresponds to $\epsilon \approx 5$.
We see that for this strain, $\eta_e$ increases strongly with the Guar gum concentration up to a value of around 0.15 \%, where the elongational viscosity starts to saturate.
These measurement are nicely correlated to the observations made in Fig. \ref{fig:EffetConcentrationGuar}.

Thus, the addition of Guar gum allows to stabilize the bubbles during generation. This was already pointed out by Frazier \textit{et al.} \cite{frazier2020make}, who showed that the elongational rheology must be taken into account to understand the choice of the artists' solutions. They have also showed how polydispersity in molecular weight of the solvated polymers leads to better performance at low concentration.





\subsection{Effect of the addition of glycerol}

\begin{figure}
\centering
\includegraphics[width=1\linewidth]{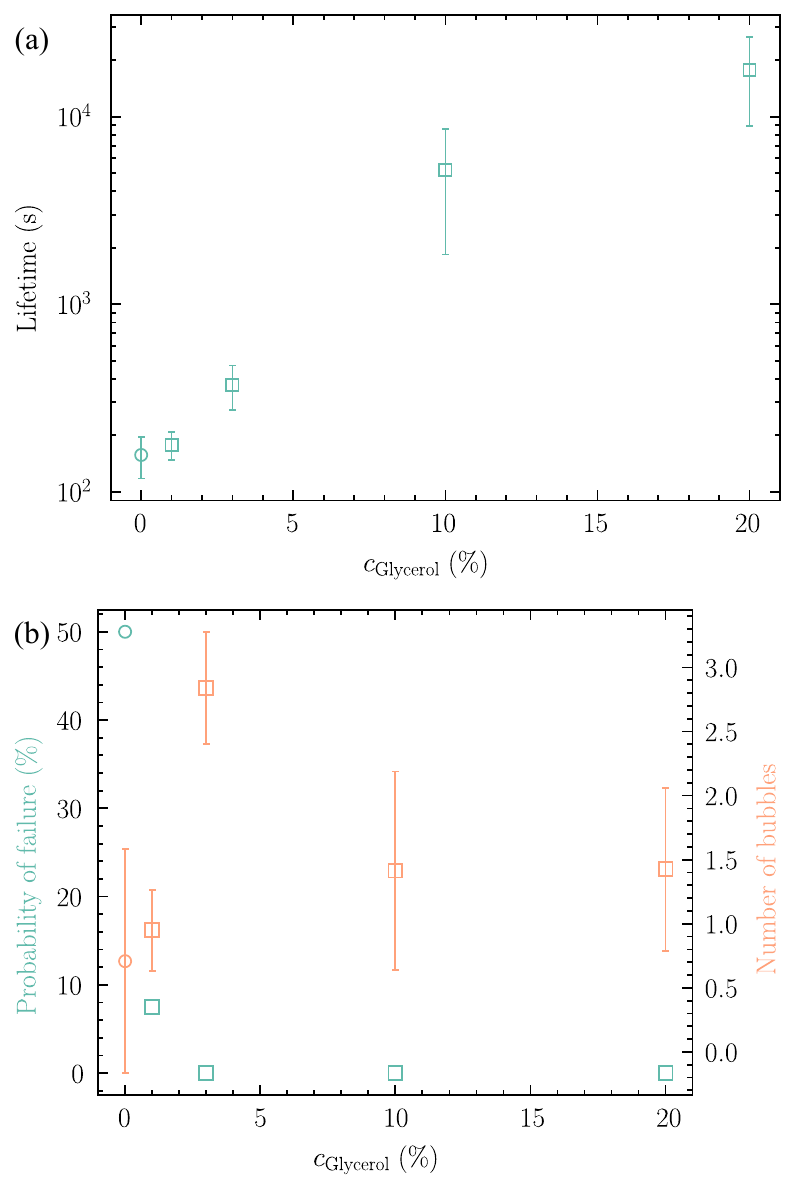}
\caption{(a) Average bubble lifetime measured as a function of the concentration of glycerol in the solution (composed with water, Fairy at a concentration of 4 \% and J-Lube at a concentration of 0.05 \%), using the device shown in Fig. \ref{fig:ManipGrosseBulle}. The scale is semi-logarithmic. The error bars correspond to the standard deviation measured on more than 10 bubbles. (b) For the same solutions, this graph shows the evolution of the failure probability (left axis) and the number of blown bubbles (right axis) obtained with the setup presented in Fig. \ref{fig:SetUpGeneration}. The error bars correspond to the standard deviation measured on about 50 trials.}
\label{fig:EffetGlycerol}
\end{figure}

We also measured the impact of adding glycerol although the ingredient is not always used by our collaborating artist, as this ingredient is found in many recipes. 
From the results presented in Fig. \ref{fig:EffetGlycerol}, the reason is obvious. 
Indeed, we gain a factor 140 on the lifetime without losing the generation efficiency, with a glycerol concentration of 20 \%.

The main mechanisms leading to film rupture is the thinning through drainage and evaporation.
Evaporation is directly linked to the atmospheric humidity and the thinning rate due to evaporation is of the order of 10-50 nm/s \cite{Miguet2020}. 
When the film is thick, it drains fast and evaporation is negligible. 
However, when the film is thin, drainage starts to be negligible and evaporation matters \cite{Champougny2018,Poulain2018a}. 
If evaporation slows down, the stability can thus increase drastically. 
Glycerol is a hygroscospic molecule \cite{Cheng2008}. 
In particular, it is known that vapor pressure of a soap film, at equilibrium depends on the film composition, that we consider to be dominated by the water-glycerol ratio.
The saturated pressure of a water-glycerol mixture can be related to \cite{D2SM00157H}: 
\begin{equation}
    p_{{\rm sat}} (x_\text{g}, T) =  p_{{\rm sat}}^0(T)  \dfrac{1 - x_\text{g}}{1 + x_\text{g} (a-1) },
    \label{eq:psatGly}
\end{equation}
\noindent
where $a = 0.248$ \cite{Glycerine1963} and $x_\text{g} = m_{\rm g} / (m_{\rm g} + m_{\rm w})$ where $m$ is a weight and the subscripts $_{\rm g}$ and $_{\rm w}$  stands for  glycerol and water, respectively.
The saturated pressure is therefore a decreasing function of the glycerol concentration. 
This is why, in the presence of glycerol, the lifetime of bubbles increases drastically \cite{Roux2021,Pasquet2022}.

\subsection{Effect of the additives to alter the pH}

\begin{figure}
\centering
\includegraphics[width=1\linewidth]{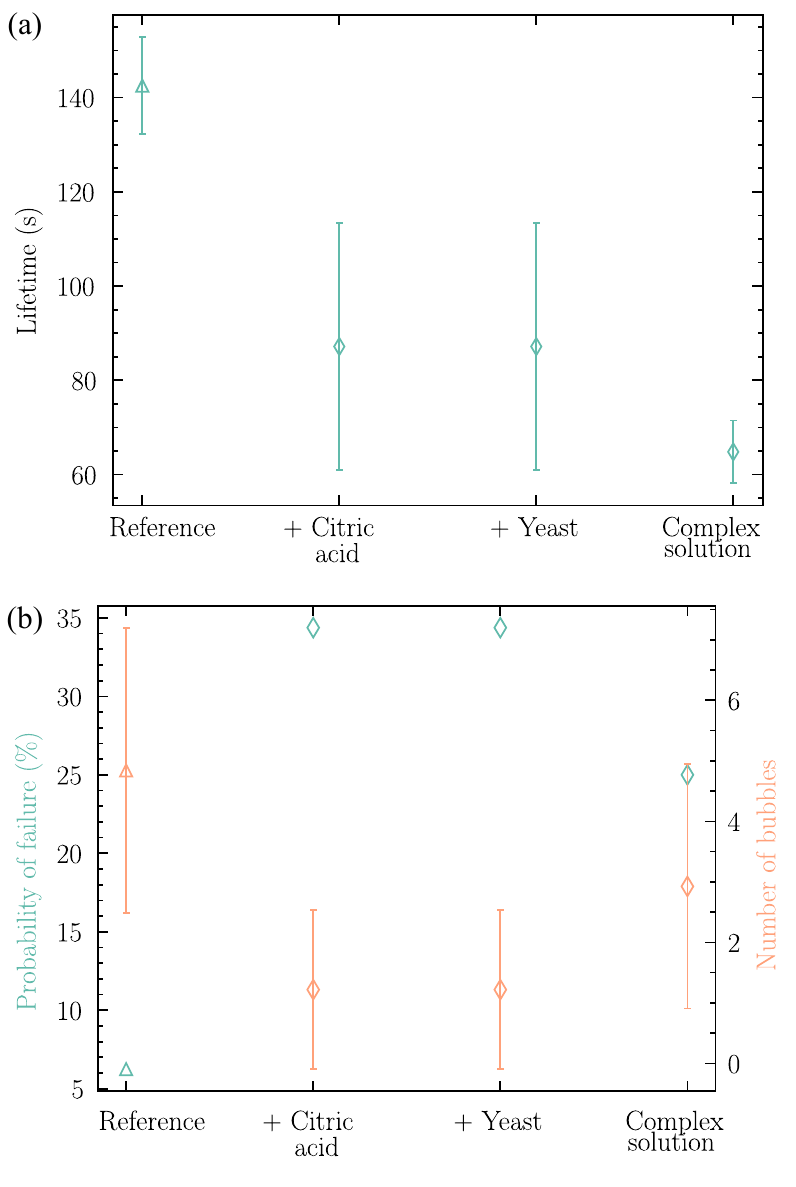}
\caption{(a) Average bubble lifetime measured using the device shown in Fig. \ref{fig:ManipGrosseBulle}. The reference solution corresponds here to the solution made with Fairy at 4 \% and J-Lube at 0.05 \%. The lifetime of the reference solution is given here together the lifetime with the reference solution with the addition of citric acid (at a concentration of 0.06 \%) or with the addition of yeast (at a concentration of 0.08 \%), respectively noted "+ Citrid acid" and "+ Yeast". The complex solution corresponds to the recipe of an artist with whom we have discussed with a concentration of Fairy at 4 \%, a concentration of J-Lube at 0.01 \%, a concentration of Guar gum at 0.1 \%, a concentration of yeast at 0.05 \% and a concentration of citric acid at 0.07 \%. The error bars correspond to the standard deviation measured on about 50 bubbles. (b) For the same solutions, this graph shows the evolution of the failure probability (left axis) and the number of blown bubbles (right axis) obtained with the setup presented in Fig. \ref{fig:SetUpGeneration}. The error bars correspond to the standard deviation measured on about 50 trials.}
\label{fig:EffectAdditives}
\end{figure}

The results obtained by adding small additives are plotted in Fig. \ref{fig:EffectAdditives}. 
The solution noted in Fig. \ref{fig:EffectAdditives} \textit{+Yeast} (pH = 8) corresponds to the recipe of our collaborating artist, the one noted \text{Complex solution} (pH = 7) to another artist we discussed with. The solution noted \textit{+Citric Acid} corresponds to the addition of citric acid to the \textit{Reference} (pH = 9) solution to obtain a pH = 7. Their composition are described in detail in the legend of Fig. \ref{fig:EffectAdditives}.
The observed effect of the additives is small and decreases both the lifetime of the bubbles and the ease with which they are generated.
It was not expected, as in the majority of recipes the artists use yeast or citric acid in the quantities studied here.

While the addition of these additives does not seem to increase the stability of the bubbles, according to some artists, it would seem that they help preserve the solutions over time. One can also think that, as they work a lot with their hands, it is more interesting for them to manipulate solutions with a pH around 7. For physics experiment, where plunging the hand in the experiments is less common, and conserving the solutions is not advised, it seems that adding these pH additives will complexify the solutions, whithout affecting neither the stability of the films nor the ease of generation.

\section{Conclusion}

Finally, we have developed controlled experiments to assess the importance of the different ingredients in a recipe proposed by the artist community
This allows us to draw some main conclusions to identify a good recipe to use in scientific studies on soap films and bubbles:
\begin{itemize}
   \item A small quantity of surfactants is necessary to allow Marangoni stress to hold the film. 4 \% in volume of Fairy dishwashing liquid seems to be a good compromise between stability and generation easiness.
   \item The addition of a lubricant with a high elongational viscosity allows an efficient bubbles generation. The concentration is not too critical and we propose a concentration of 0.1 \% of Guar gum. For less controlled experiments, J-Lube can be used at a concentration of 0.05 \%. 
   \item The addition of  glycerol at a concentration of 10 \% allows to slow evaporation down and thus to increase the bubble lifetime
   \item The presence of additives to alter the pH seems inefficient. However, they may help to conserve the solutions.
\end{itemize}
Of course, we did not provide here any proof that this recipe will be the best possible recipe, but it is based on systematic studies and clear physical arguments. We hope that it could help, to some points, to create a standard for the scientific community.

\section*{Author contribution statement}
M.P. and L.W. carried out the experiment. M.P., F.R. and E.R. wrote the manuscript. F.R. and E.R. supervised the project. P-Y.F. helped conceive the idea of the experiments. All authors provided critical feedback and helped shape the research, analysis and manuscript.

\section*{Data Availability Statement}
The datasets generated and analysed during the current study are available from the corresponding author on reasonable request.

\section*{Acknowledgments}
We thank Jérémie Sanchez for the help he gave us and for the development of the LabView interface. We also thank Serge Guichard for the nice pictures. 
We are grateful to all the members, both artists and researchers of the workshop "Faire des bulles, tout un art, toute une science" and in particular to the artists, who shared their recipes with us. 
Funding from ESA (MAP Soft Matter Dynamics) ANR (ANR-20-CE30-0019) and CNES (through the GDR MFA) is acknowledged.


\bibliography{references} 

\providecommand*{\mcitethebibliography}{\thebibliography}
\csname @ifundefined\endcsname{endmcitethebibliography}
{\let\endmcitethebibliography\endthebibliography}{}
\begin{mcitethebibliography}{32}
\providecommand*{\natexlab}[1]{#1}
\providecommand*{\mciteSetBstSublistMode}[1]{}
\providecommand*{\mciteSetBstMaxWidthForm}[2]{}
\providecommand*{\mciteBstWouldAddEndPuncttrue}
  {\def\EndOfBibitem{\unskip.}}
\providecommand*{\mciteBstWouldAddEndPunctfalse}
  {\let\EndOfBibitem\relax}
\providecommand*{\mciteSetBstMidEndSepPunct}[3]{}
\providecommand*{\mciteSetBstSublistLabelBeginEnd}[3]{}
\providecommand*{\EndOfBibitem}{}
\mciteSetBstSublistMode{f}
\mciteSetBstMaxWidthForm{subitem}
{(\emph{\alph{mcitesubitemcount}})}
\mciteSetBstSublistLabelBeginEnd{\mcitemaxwidthsubitemform\space}
{\relax}{\relax}

\bibitem[Hooke(1672)]{hooke28holes}
R.~Hooke, \emph{Communications to the Royal Society}, 1672\relax
\mciteBstWouldAddEndPuncttrue
\mciteSetBstMidEndSepPunct{\mcitedefaultmidpunct}
{\mcitedefaultendpunct}{\mcitedefaultseppunct}\relax
\EndOfBibitem
\bibitem[Newton(1952)]{newton1952opticks}
I.~Newton, \emph{New York: Dover Publications Inc}, 1952\relax
\mciteBstWouldAddEndPuncttrue
\mciteSetBstMidEndSepPunct{\mcitedefaultmidpunct}
{\mcitedefaultendpunct}{\mcitedefaultseppunct}\relax
\EndOfBibitem
\bibitem[Boys(1888)]{boys1888li}
C.~V. Boys, \emph{The London, Edinburgh, and Dublin Philosophical Magazine and
  Journal of Science}, 1888, \textbf{25}, 409--419\relax
\mciteBstWouldAddEndPuncttrue
\mciteSetBstMidEndSepPunct{\mcitedefaultmidpunct}
{\mcitedefaultendpunct}{\mcitedefaultseppunct}\relax
\EndOfBibitem
\bibitem[Isenberg(1992)]{isenberg1992science}
C.~Isenberg, \emph{The science of soap films and soap bubbles}, Courier Dover
  Publications, 1992\relax
\mciteBstWouldAddEndPuncttrue
\mciteSetBstMidEndSepPunct{\mcitedefaultmidpunct}
{\mcitedefaultendpunct}{\mcitedefaultseppunct}\relax
\EndOfBibitem
\bibitem[Weaire and Drenckhan(2008)]{WEAIRE200820}
D.~Weaire and W.~Drenckhan, \emph{Advances in Colloid and Interface Science},
  2008, \textbf{137}, 20--26\relax
\mciteBstWouldAddEndPuncttrue
\mciteSetBstMidEndSepPunct{\mcitedefaultmidpunct}
{\mcitedefaultendpunct}{\mcitedefaultseppunct}\relax
\EndOfBibitem
\bibitem[Drenckhan and Hutzler(2015)]{drenckhan2015structure}
W.~Drenckhan and S.~Hutzler, \emph{Advances in colloid and interface science},
  2015, \textbf{224}, 1--16\relax
\mciteBstWouldAddEndPuncttrue
\mciteSetBstMidEndSepPunct{\mcitedefaultmidpunct}
{\mcitedefaultendpunct}{\mcitedefaultseppunct}\relax
\EndOfBibitem
\bibitem[Kellay and Goldburg(2002)]{Kellay_2002}
H.~Kellay and W.~I. Goldburg, \emph{Reports on Progress in Physics}, 2002,
  \textbf{65}, 845--894\relax
\mciteBstWouldAddEndPuncttrue
\mciteSetBstMidEndSepPunct{\mcitedefaultmidpunct}
{\mcitedefaultendpunct}{\mcitedefaultseppunct}\relax
\EndOfBibitem
\bibitem[Chomaz and Cathalau(1990)]{chomaz1990soap}
J.~Chomaz and B.~Cathalau, \emph{Physical Review A}, 1990, \textbf{41},
  2243\relax
\mciteBstWouldAddEndPuncttrue
\mciteSetBstMidEndSepPunct{\mcitedefaultmidpunct}
{\mcitedefaultendpunct}{\mcitedefaultseppunct}\relax
\EndOfBibitem
\bibitem[Couder \emph{et~al.}(1989)Couder, Chomaz, and
  Rabaud]{couder1989hydrodynamics}
Y.~Couder, J.~Chomaz and M.~Rabaud, \emph{Physica D: Nonlinear Phenomena},
  1989, \textbf{37}, 384--405\relax
\mciteBstWouldAddEndPuncttrue
\mciteSetBstMidEndSepPunct{\mcitedefaultmidpunct}
{\mcitedefaultendpunct}{\mcitedefaultseppunct}\relax
\EndOfBibitem
\bibitem[Kellay(2017)]{Kellay_POF_2017}
H.~Kellay, \emph{Physics of Fluids}, 2017, \textbf{29}, 111113\relax
\mciteBstWouldAddEndPuncttrue
\mciteSetBstMidEndSepPunct{\mcitedefaultmidpunct}
{\mcitedefaultendpunct}{\mcitedefaultseppunct}\relax
\EndOfBibitem
\bibitem[{Guinness World Record}(2015)]{webPealman}
{Guinness World Record}, \emph{Largest free floating soap bubble (outdoors)},
  2015,
  \url{https://www.guinnessworldrecords.com/world-records/largest-free-floating-soap-bubble}\relax
\mciteBstWouldAddEndPuncttrue
\mciteSetBstMidEndSepPunct{\mcitedefaultmidpunct}
{\mcitedefaultendpunct}{\mcitedefaultseppunct}\relax
\EndOfBibitem
\bibitem[Soa()]{Soapwiki}
\emph{Soap Bubble Wiki},
  \url{https://soapbubble.fandom.com/wiki/Soap_Bubble_Wiki}\relax
\mciteBstWouldAddEndPuncttrue
\mciteSetBstMidEndSepPunct{\mcitedefaultmidpunct}
{\mcitedefaultendpunct}{\mcitedefaultseppunct}\relax
\EndOfBibitem
\bibitem[Frazier \emph{et~al.}(2020)Frazier, Jiang, and
  Burton]{frazier2020make}
S.~Frazier, X.~Jiang and J.~C. Burton, \emph{Physical Review Fluids}, 2020,
  \textbf{5}, 013304\relax
\mciteBstWouldAddEndPuncttrue
\mciteSetBstMidEndSepPunct{\mcitedefaultmidpunct}
{\mcitedefaultendpunct}{\mcitedefaultseppunct}\relax
\EndOfBibitem
\bibitem[Hamlett \emph{et~al.}(2021)Hamlett, Boniface, Salonen, Rio, Perkins,
  Clark, Nyugen, and Fairhurst]{hamlett2021blowing}
C.~A. Hamlett, D.~N. Boniface, A.~Salonen, E.~Rio, C.~Perkins, A.~Clark,
  S.~Nyugen and D.~J. Fairhurst, \emph{Soft Matter}, 2021, \textbf{17},
  2404--2409\relax
\mciteBstWouldAddEndPuncttrue
\mciteSetBstMidEndSepPunct{\mcitedefaultmidpunct}
{\mcitedefaultendpunct}{\mcitedefaultseppunct}\relax
\EndOfBibitem
\bibitem[Salkin \emph{et~al.}(2016)Salkin, Schmit, Panizza, and
  Courbin]{salkin2016generating}
L.~Salkin, A.~Schmit, P.~Panizza and L.~Courbin, \emph{Physical review
  letters}, 2016, \textbf{116}, 077801\relax
\mciteBstWouldAddEndPuncttrue
\mciteSetBstMidEndSepPunct{\mcitedefaultmidpunct}
{\mcitedefaultendpunct}{\mcitedefaultseppunct}\relax
\EndOfBibitem
\bibitem[Boulogne(2019)]{Boulogne2019}
F.~Boulogne, \emph{The European Physical Journal E}, 2019, \textbf{42},
  51\relax
\mciteBstWouldAddEndPuncttrue
\mciteSetBstMidEndSepPunct{\mcitedefaultmidpunct}
{\mcitedefaultendpunct}{\mcitedefaultseppunct}\relax
\EndOfBibitem
\bibitem[Cohen \emph{et~al.}(2017)Cohen, Texier, Reyssat, Snoeijer,
  Qu{\'e}r{\'e}, and Clanet]{cohen2017shape}
C.~Cohen, B.~D. Texier, E.~Reyssat, J.~H. Snoeijer, D.~Qu{\'e}r{\'e} and
  C.~Clanet, \emph{Proceedings of the National Academy of Sciences}, 2017,
  \textbf{114}, 2515--2519\relax
\mciteBstWouldAddEndPuncttrue
\mciteSetBstMidEndSepPunct{\mcitedefaultmidpunct}
{\mcitedefaultendpunct}{\mcitedefaultseppunct}\relax
\EndOfBibitem
\bibitem[Dinic \emph{et~al.}(2017)Dinic, Biagioli, and Sharma]{Dinic2017}
J.~Dinic, M.~Biagioli and V.~Sharma, \emph{Journal of Polymer Science Part B:
  Polymer Physics}, 2017, \textbf{55}, 1692--1704\relax
\mciteBstWouldAddEndPuncttrue
\mciteSetBstMidEndSepPunct{\mcitedefaultmidpunct}
{\mcitedefaultendpunct}{\mcitedefaultseppunct}\relax
\EndOfBibitem
\bibitem[Ballet and Graner(2006)]{ballet2006giant}
P.~Ballet and F.~Graner, \emph{European journal of physics}, 2006, \textbf{27},
  951\relax
\mciteBstWouldAddEndPuncttrue
\mciteSetBstMidEndSepPunct{\mcitedefaultmidpunct}
{\mcitedefaultendpunct}{\mcitedefaultseppunct}\relax
\EndOfBibitem
\bibitem[Bergeron(1997)]{Bergeron1995}
V.~Bergeron, \emph{Langmuir}, 1997, \textbf{13}, 3474--3482\relax
\mciteBstWouldAddEndPuncttrue
\mciteSetBstMidEndSepPunct{\mcitedefaultmidpunct}
{\mcitedefaultendpunct}{\mcitedefaultseppunct}\relax
\EndOfBibitem
\bibitem[Lin \emph{et~al.}(1999)Lin, Lin, Chen, Hsu, and Kwan]{Lin1999}
S.-Y. Lin, Y.-Y. Lin, E.-M. Chen, C.-T. Hsu and C.-C. Kwan, \emph{Langmuir},
  1999, \textbf{15}, 4370--4376\relax
\mciteBstWouldAddEndPuncttrue
\mciteSetBstMidEndSepPunct{\mcitedefaultmidpunct}
{\mcitedefaultendpunct}{\mcitedefaultseppunct}\relax
\EndOfBibitem
\bibitem[Mysels(1986)]{mysels1986surface}
K.~J. Mysels, \emph{Langmuir}, 1986, \textbf{2}, 423--428\relax
\mciteBstWouldAddEndPuncttrue
\mciteSetBstMidEndSepPunct{\mcitedefaultmidpunct}
{\mcitedefaultendpunct}{\mcitedefaultseppunct}\relax
\EndOfBibitem
\bibitem[de~Gennes(2001)]{DeGennes_2001}
P.~G. de~Gennes, \emph{Langmuir}, 2001, \textbf{17}, 2416--2419\relax
\mciteBstWouldAddEndPuncttrue
\mciteSetBstMidEndSepPunct{\mcitedefaultmidpunct}
{\mcitedefaultendpunct}{\mcitedefaultseppunct}\relax
\EndOfBibitem
\bibitem[Lucassen-Reynders(1981)]{Lucassen1981}
E.~Lucassen-Reynders, in \emph{Anionic surfactants: Physical chemistry of
  surfactant action}, M. Dekker, 1981, ch.~6\relax
\mciteBstWouldAddEndPuncttrue
\mciteSetBstMidEndSepPunct{\mcitedefaultmidpunct}
{\mcitedefaultendpunct}{\mcitedefaultseppunct}\relax
\EndOfBibitem
\bibitem[Miguet \emph{et~al.}(2020)Miguet, Pasquet, Rouyer, Fang, and
  Rio]{Miguet2020}
J.~Miguet, M.~Pasquet, F.~Rouyer, Y.~Fang and E.~Rio, \emph{Soft Matter}, 2020,
  \textbf{16}, 1082--1090\relax
\mciteBstWouldAddEndPuncttrue
\mciteSetBstMidEndSepPunct{\mcitedefaultmidpunct}
{\mcitedefaultendpunct}{\mcitedefaultseppunct}\relax
\EndOfBibitem
\bibitem[Champougny \emph{et~al.}(2018)Champougny, Miguet, Henaff, Restagno,
  Boulogne, and Rio]{Champougny2018}
L.~Champougny, J.~Miguet, R.~Henaff, F.~Restagno, F.~Boulogne and E.~Rio,
  \emph{Langmuir}, 2018, \textbf{34}, 3221--3227\relax
\mciteBstWouldAddEndPuncttrue
\mciteSetBstMidEndSepPunct{\mcitedefaultmidpunct}
{\mcitedefaultendpunct}{\mcitedefaultseppunct}\relax
\EndOfBibitem
\bibitem[Poulain and Bourouiba(2018)]{Poulain2018a}
S.~Poulain and L.~Bourouiba, \emph{Physical Review Letters}, 2018,
  \textbf{121}, 204502\relax
\mciteBstWouldAddEndPuncttrue
\mciteSetBstMidEndSepPunct{\mcitedefaultmidpunct}
{\mcitedefaultendpunct}{\mcitedefaultseppunct}\relax
\EndOfBibitem
\bibitem[Cheng(2008)]{Cheng2008}
N.-S. Cheng, \emph{Industrial \& Engineering Chemistry Research}, 2008,
  \textbf{47}, 3285--3288\relax
\mciteBstWouldAddEndPuncttrue
\mciteSetBstMidEndSepPunct{\mcitedefaultmidpunct}
{\mcitedefaultendpunct}{\mcitedefaultseppunct}\relax
\EndOfBibitem
\bibitem[Pasquet \emph{et~al.}(2022)Pasquet, Boulogne, Sant-Anna, Restagno, and
  Rio]{D2SM00157H}
M.~Pasquet, F.~Boulogne, J.~Sant-Anna, F.~Restagno and E.~Rio, \emph{Soft
  Matter}, 2022, \textbf{18}, 4536--4542\relax
\mciteBstWouldAddEndPuncttrue
\mciteSetBstMidEndSepPunct{\mcitedefaultmidpunct}
{\mcitedefaultendpunct}{\mcitedefaultseppunct}\relax
\EndOfBibitem
\bibitem[Association
  \emph{et~al.}(1963)Association\emph{et~al.}]{Glycerine1963}
G.~P. Association \emph{et~al.}, \emph{Physical properties of glycerine and its
  solutions}, Glycerine Producers' Association, 1963\relax
\mciteBstWouldAddEndPuncttrue
\mciteSetBstMidEndSepPunct{\mcitedefaultmidpunct}
{\mcitedefaultendpunct}{\mcitedefaultseppunct}\relax
\EndOfBibitem
\bibitem[Roux \emph{et~al.}(2022)Roux, Duchesne, and Baudoin]{Roux2021}
A.~Roux, A.~Duchesne and M.~Baudoin, \emph{Phys. Rev. Fluids}, 2022,
  \textbf{7}, L011601\relax
\mciteBstWouldAddEndPuncttrue
\mciteSetBstMidEndSepPunct{\mcitedefaultmidpunct}
{\mcitedefaultendpunct}{\mcitedefaultseppunct}\relax
\EndOfBibitem
\bibitem[Pasquet \emph{et~al.}(2022)Pasquet, Boulogne, Sant-Anna, Restagno, and
  Rio]{Pasquet2022}
M.~Pasquet, F.~Boulogne, J.~Sant-Anna, F.~Restagno and E.~Rio, \emph{Soft
  Matter}, 2022, \textbf{18}, 4536--4542\relax
\mciteBstWouldAddEndPuncttrue
\mciteSetBstMidEndSepPunct{\mcitedefaultmidpunct}
{\mcitedefaultendpunct}{\mcitedefaultseppunct}\relax
\EndOfBibitem
\end{mcitethebibliography}
\bibliographystyle{rsc} 
\end{document}